\documentclass[twocolumn,aps,prl,superscriptaddress,amsmath]{revtex4-2}
\usepackage[utf8]{inputenc}
\usepackage[lcgreekalpha]{stix2}
\usepackage{bm}
\usepackage{xcolor}
\usepackage{graphicx}

\begin{document}

\title{Higher Landau-Level Analogs and \\ Signatures of Non-Abelian States in Twisted Bilayer MoTe$_2$}

\author{Chong Wang}
\thanks{These authors contribute equally to the work.}
\affiliation{Department of Materials Science and Engineering, University of Washington, Seattle, WA 98195, USA}
\author{Xiao-Wei Zhang}
\thanks{These authors contribute equally to the work.}
\affiliation{Department of Materials Science and Engineering, University of Washington, Seattle, WA 98195, USA}
\author{Xiaoyu Liu}
\affiliation{Department of Materials Science and Engineering, University of Washington, Seattle, WA 98195, USA}
\author{Jie Wang}
\affiliation{Department of Physics, Temple University, Philadelphia, Pennsylvania, 19122, USA}
\author{Ting Cao}
\email{tingcao@uw.edu}
\affiliation{Department of Materials Science and Engineering, University of Washington, Seattle, WA 98195, USA}
\author{Di Xiao}
\email{dixiao@uw.edu}
\affiliation{Department of Materials Science and Engineering, University of Washington, Seattle, WA 98195, USA}
\affiliation{Department of Physics, University of Washington, Seattle, WA 98195, USA}

\begin{abstract}
Recent experimental discovery of fractional Chern insulators at zero magnetic field in moir\'e superlattices has sparked intense interests in bringing Landau level physics to flat Chern bands. In twisted MoTe$_2$ bilayers (tMoTe$_2$), recent theoretical and experimental studies have found three consecutive flat Chern bands at twist angle $\sim 2^\circ$. In this work, we investigate whether higher Landau level physics can be found in these consecutive Chern bands. At twist angles $2.00^\circ$ and $1.89^\circ$, we identify four consecutive $C = 1$ bands for the $K$ valley in tMoTe$_2$. By constructing Wannier functions directly from density functional theory (DFT) calculations, a six-orbital model is developed to describe the consecutive Chern bands, with the orbitals forming a honeycomb lattice. Exact diagonalization on top of Hartree-Fock calculations are carried out with the Wannier functions. Especially, when the second moir\'e miniband is half-filled, signatures of non-Abelian states are found. Our Wannier-based approach in modelling moir\'e superlattices is faithful to DFT wave functions and can serve as benchmarks for continuum models. The possibility of realizing non-Abelian anyons at zero magnetic field also opens up a new pathway for fault-tolerant quantum information processing.
\end{abstract}

\maketitle 

\textit{Introduction}.---Recent experiments~\cite{cai2023signatures,park2023observation,zeng2023thermodynamic,PhysRevX.13.031037,lu2024fractional} and theories~\cite{PhysRevResearch.3.L032070,PhysRevLett.132.036501,PhysRevB.108.085117,PhysRevB.107.L201109,PhysRevResearch.5.L032022,PhysRevLett.131.136502,PhysRevLett.131.136501} have identified a series of Abelian fractional Chern insulators (FCI), such as the Laughlin states and other Jain sequence states, as well as gapless composite fermi liquids (CFL), in moir\'e superlattices at zero magnetic field. The emergence of these exotic states is attributed to the existence of flat Chern bands~\cite{PhysRevLett.106.236804,PhysRevLett.106.236802,sheng2011fractional,PhysRevX.1.021014} in these systems. Specifically, in twisted homobilayer transition metal dichalcogenides, within the framework of the continuum model, electrons can be viewed as hopping in a layer pseudospin skyrmion lattice, giving rise to topologically nontrivial flat bands~\cite{wu_topological_2019,yu2020giant,PhysRevLett.132.096602}. The quantum geometry of these flat Chern bands resembling that of the lowest Landau level (LLL) is one of the important reasons for the emergence of the above exotic states~\cite{PhysRevB.85.241308,PhysRevB.90.165139,kahlerband1,kahlerband2,PhysRevLett.127.246403,ledwith2022vortexability,Jie_GMP,HigherVortexability24}.

More exotic states, such as the Moore-Read (MR) state featuring non-Abelian excitations~\cite{moore1991nonabelions}, can be stabilized by Coulomb interaction in the first Landau Level (LL)~\cite{PhysRevLett.59.1776,PhysRevLett.104.076803,PhysRevLett.96.016803,PhysRevLett.104.086801}. In a recent density functional theory (DFT) study of twisted bilayer MoTe$_2$ (tMoTe$_2$), it is discovered that three consecutive flat bands with Chern numbers equal to 1 appear at twist angle $2.14^\circ$ for each valley~\cite{zhang2024polarization}, which has not been predicted in the continuum model description within the first harmonic approximation. The existence of three consecutive flat Chern bands has also been corroborated by experimental observations at similar twist angles~\cite{kang2024evidence}. These consecutive flat Chern bands with identical Chern numbers bear a striking resemblance to the series of LLs, hinting at the possibility of bringing higher LL physics to moir\'e superlattices at zero magnetic field, especially the non-Abelian state such as MR.

In this work, we first extend our previous large-scale DFT calculations on tMoTe$_2$ to additional twist angles near $2^\circ$. We identify four consecutive $C=1$ bands at twist angles $2.00^\circ$ and $1.89^\circ$ for the $K$ valley. To accurately describe the quantum geometry, we construct Wannier functions directly from DFT calculations and develop a six-orbital model to describe the consecutive Chern bands, where the orbitals form a honeycomb lattice. For the second moir\'e miniband, the integral of the trace of the Fubini-Study metric [$\mathrm{tr}(g)$] is shown to be close to that of the first LL. In addition, the fluctuations of Berry curvature and $\mathrm{tr}(g)$ can be significantly suppressed by band mixing in Hartree-Fock (HF) calculations, enhancing the analogy between the second moir\'e miniband and the first LL. Exact diagonalization (ED) on top of HF calculations are carried out using the Wannier functions. When the second moir\'e miniband is half-filled, signatures of non-Abelian states are found. Our Wannier-based approach in modelling the moir\'e superlattice is faithful to DFT wave functions and can serve as benchmarks for continuum models. The possibility of realizing non-Abelian states in tMoTe$_2$ also opens up an exciting avenue in realizing higher LL physics in moir\'e superlattices.

\textit{Consecutive Chern bands in tMoTe$_2$}.---
When moir\'e superlattices are formed by twisting two identical monolayers, the original monolayer bands are broken into minibands. For monolayer MoTe$_2$, the valence band top is located at two corners ($K$ and $K'$) of the honeycomb Brillouin zone. In moir\'e superlattices, the bands from both $K$ and $K'$ valley form two independent sets of minibands. The two sets of bands are partners under the operation of time reversal symmetry.
Our previous large-scale density-functional-theory (DFT) calculations with machine learning force fields has revealed an intricate dependence of the band topology on twist angles~\cite{zhang2024polarization}. Interestingly, at twist angle of $2.14^\circ$, $K$ valley valence bands of tMoTe$_2$ feature three consecutive Chern bands with $C=1$. With hole doping, these Chern bands can be revealed in experiments when valleys are spontaneously polarized due to electron-electron interactions.

\begin{figure}
\centering
\includegraphics[width=0.927\columnwidth]{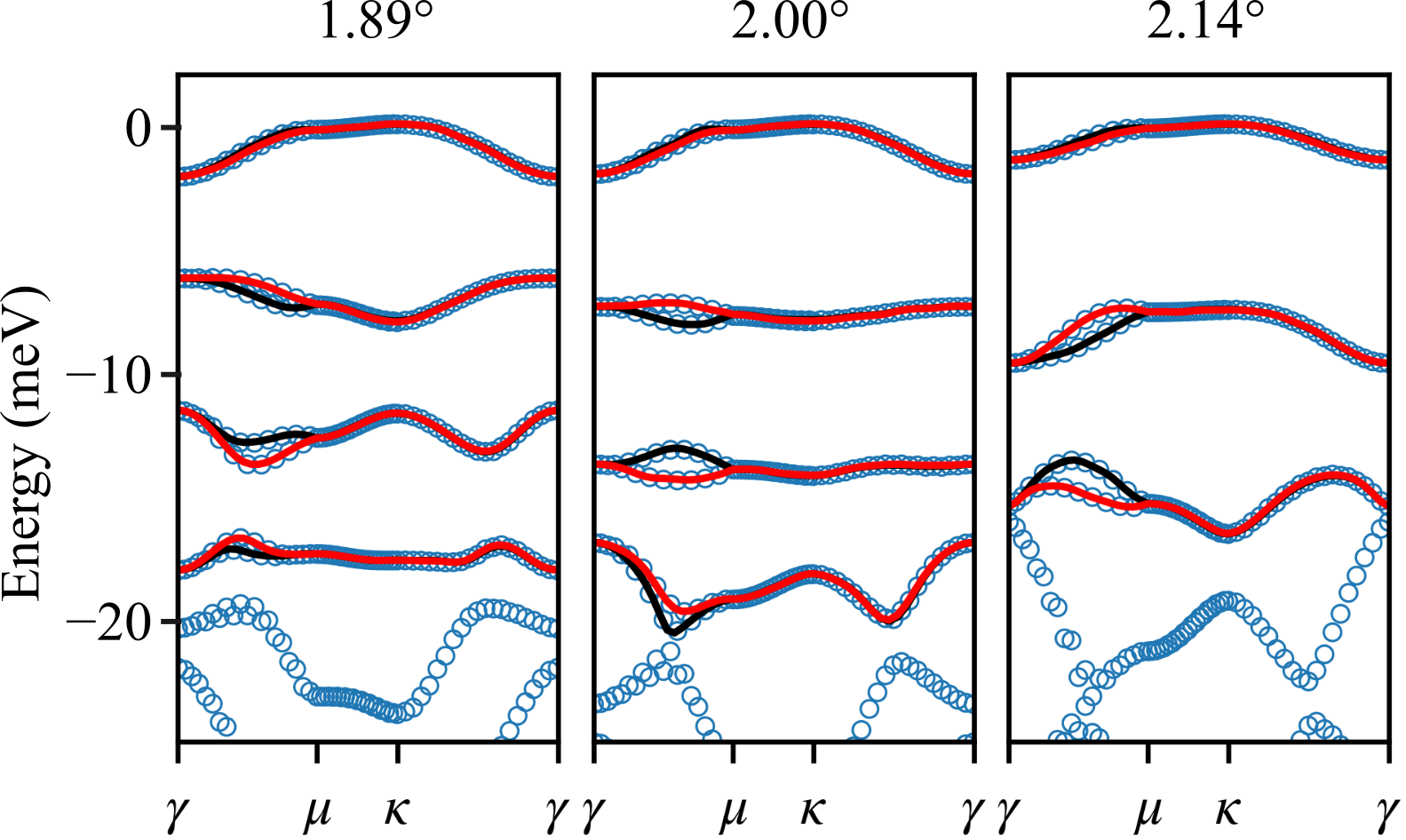}
\caption{
Bands for tMoTe$_2$ at twist angles $1.89^\circ$, $2.00^\circ$ and $2.14^\circ$. Empty blue circles are from DFT calculations, and solid lines are from Wannier interpolation. For Wannier interpolated bands, red lines are from the $K$ (spin up) valley and black lines are from the $K'$ (spin down) valley. All Wannier interpolated bands have Chern number $C = +1$ for the $K$ valley. Only Wannier interpolated bands in the frozen window are shown here.
\label{fig:bands}}
\end{figure}

To further explore the band topology of tMoTe$_2$ around this twist angle, we perform DFT calculations at two other twist angles $2.00^\circ$ and $1.89^\circ$, following the same method introduced in Ref.~[\onlinecite{zhang2024polarization}]{, where the DFT-D2 Van der Waals correction~\cite{grimme2006semiempirical} is used~\cite{zhang2024polarization,PhysRevB.109.205121}}.  The moir\'e valence bands from all three angles are shown in Fig.~\ref{fig:bands}.  Four consecutive $C=1$ bands from $K$ valley are found at twist angles $2.00^\circ$ and $1.89^\circ$.  At twist angle $2.14^\circ$, the fourth band is not well-isolated to have a well-defined Chern number.  The Chern numbers are determined by Wannier interpolation with Wannier functions, whose construction will be described below.

The consecutive Chern bands are flat, hinting at the possibility of strong-correlated physics when these bands are partially filled. In this work, we focus on the possibility of realizing first LL physics in tMoTe$_2$, specifically targeting the second valence band (bands are numbered in the descending order of energy). At the twist angle $2.00^\circ$, the second valence band reaches optimal flatness. Therefore, we will focus on this twist angle in the following, deferring the results from other twist angles to Supplemental Material~\cite{supplemental}.

\begin{figure}
\centering
\includegraphics[width=0.991\columnwidth]{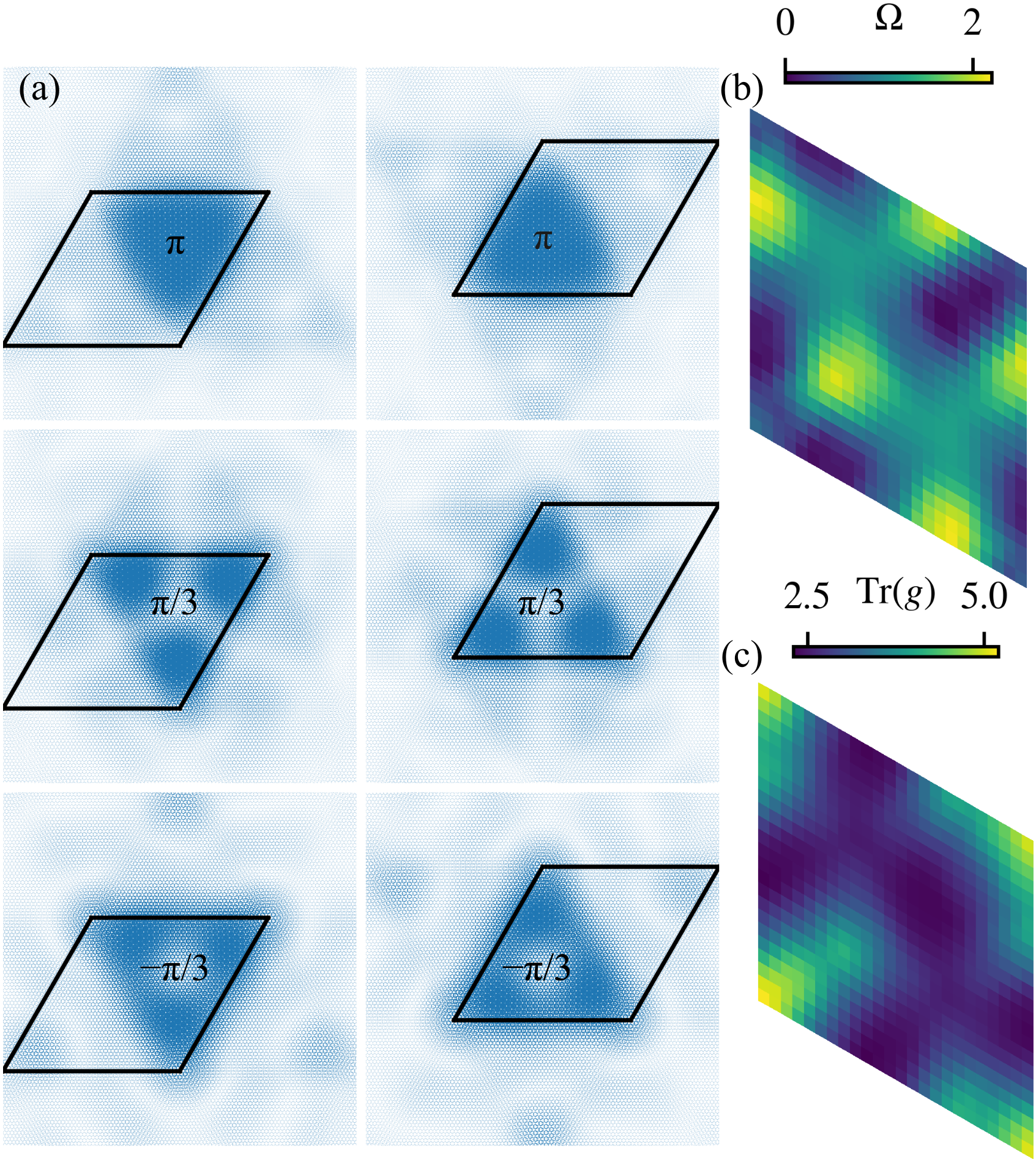}
\caption{
(a) Real space distributions of the Wannier functions for tMoTe$_2$ at twist angle $2.00^\circ$. Black parallelograms represent moir\'e unit cell. The phases of the $C_3$ eigenvalues with respect to the center of the Wannier functions have been labeled. Contributions from both layers have been summed over. (b) and (c) show the distribution of $\Omega$ and $\mathrm{tr}(g)$ in the Brillouin zone for the second moir\'e miniband. The unit for both $\Omega$ and $\mathrm{tr}(g)$ is $2\mathrm{\pi}/|\Gamma|$, where $|\Gamma|$ is the area of the Brillouin zone. Both $\Omega$ and $\mathrm{tr}(g)$ are calculated from the small-$\bm{q}$ expansion of the form factors.
\label{fig:wannier}}
\end{figure}

\textit{Quantum geometry from Wannier functions.}---To investigate whether consecutive moir\'e valence bands resemble LL series, accurate modelling is required to capture fluctuations of the quantum geometry of the moir\'e minibands. Currently, the most common description of moir\'e superlattice is continuum models, in which the effect of moir\'e superlattice is described by moir\'e potentials periodic in the moir\'e Bravais lattice vectors. However, continuum models fitted to DFT bands within the first few harmonic moir\'e potentials do not guarantee an accurate reproduction of the quantum geometry{~\cite{zhang2024polarization}} such as the Berry curvature $\Omega$ and the Fubini-Study metric $g$ from DFT wave functions. Here, we construct Wannier functions to faithfully represent DFT wave functions{~\cite{RevModPhys.84.1419}} and perform many-body calculations on top of the Wannier functions. The Wannier functions are constructed for the $K$ valley bands, and the $K'$ valley Wannier functions are obtained using time reversal symmetry. The valleys in the DFT calculations are decoupled by distinct Bloch phases and opposite expectation values of the spin-$z$ operator.

Our approach to construct the Wannier functions is the ``projection'' method~\cite{PhysRev.135.A685}, which is also the first step in constructing maximally localized Wannier functions~\cite{PhysRevB.56.12847,RevModPhys.84.1419}. This approach first chooses several trial Wannier functions and then projects the relevant Bloch states onto the trial Wannier functions. The projected Bloch states are subsequently orthogonalized. The Fourier transformation of the orthogonalized projected Bloch states gives the desired Wannier functions. This method generally retains the symmetry properties of the trial Wannier functions~\cite{RevModPhys.84.1419} and is a powerful tool to construct tight-binding models from DFT calculations.

The DFT bands do not possess a local gap above which the total Chern number is zero. Therefore, band disentanglement~\cite{PhysRevB.65.035109} needs to be performed to avoid Wannier obstruction~\cite{PhysRevLett.98.046402}. A set of frozen states is chosen for which the DFT band energies and Bloch states are faithfully reconstructed. For twist angle $2.14^\circ$, we choose the first three valence bands as frozen states. For twist angles $2.00^\circ$ and $1.89^\circ$, first four bands are frozen.

For twist angle $2.00^\circ$, the real space plots of the Wannier functions are shown in Fig.~\ref{fig:wannier}(a). The Wannier functions form a honeycomb lattice. For each site in the honeycomb lattice, there are three Wannier functions with different three-fold rotation symmetry [$C_3$] eigenvalues. We have chosen trial Wannier functions centered at MX (Mo on top of Te) and XM (Te on top of Mo) stackings. At some high symmetry $\bm{k}$ points, DFT wave functions for certain bands are localized at the MM (Mo on top of Mo) stacking~\cite{PhysRevX.13.041026}, which is covered by linear combinations of the Wannier functions. The Wannier-interpolated bands, along with the DFT bands, are shown in Fig.~\ref{fig:bands}, where an excellent agreement is observed.

In this work, we mainly employ two indicators to compare the quantum geometry of moir\'e minibands and LLs, namely the trace of the Fubini-Study metric $\mathrm{tr}(g)$ and the Berry curvature. For LLs, the integration of $\mathrm{tr}(g)$ 
\begin{equation}
    \chi = \frac{1}{2 \mathrm{\pi}}\int_{\mathrm{BZ}} \mathrm{d}\bm{k} \mathrm{tr}[g(\bm{k})]
\end{equation}
is $2 n + 1$, where $n$ is the LL index{~\cite{liu2024theory}}. With Wannier functions, our calculated results of $\chi$ are $1.04$, $3.09$, $5.11$, $7.53$ for the topmost four moir\'e bands, resembling that of LLs. Another important feature of LLs is that they have flat $\Omega$ and $\mathrm{tr}(g)$. Targeting the second moir\'e valence, we plot the distribution of $\Omega$ and $\mathrm{tr}(g)$ in the Brillouin zone in Fig.~\ref{fig:wannier}(b,c). The fluctuation of $\Omega$ and $\mathrm{tr}(g)$ is relatively large, with the standard deviation being 0.51 and 0.67, respectively (Table \ref{tab:angles}). We will show that electron-electron interactions can improve the flatness of $\Omega$ and $\mathrm{tr}(g)$ by band mixing.

\textit{Coulomb interaction}.---To include the electron-electron interactions, we adopt the following interacting Hamiltonian
\begin{equation}
\begin{aligned}
& \hat{H}_\text{int} = \sum_{\{n\},\{\bm{k}\},\bm{q}} V_{n_1 n_2 n_3 n_4}(\bm{k}_1,\bm{k}_2, \bm{q}) \hat{a}^\dagger_{n_1 \bm{k}_1} \hat{a}^\dagger_{n_2 \bm{k}_2} \hat{a}_{n_3 \bm{k}_2 -\bm{q}} \hat{a}_{n_4 \bm{k}_1 + \bm{q}}, \\
& V_{n_1 n_2 n_3 n_4}(\bm{k}_1,\bm{k}_2, \bm{q}) = \frac{1}{2A} v(\bm{q}) f_{n_1 n_4}(\bm{k}_1, \bm{q}) f_{n_2 n_3}(\bm{k}_2, -\bm{q}),
\end{aligned}
\end{equation}
where $\hat{a}^\dagger_{n \bm{q}}$ creates a Bloch state for band $n$ at crystal momentum $\bm{q}$ and $A$ is the area of the system. The summation over $\bm{k}$ is in the Brillouin zone and the summation of $\bm{q}$ is unbounded. We have taken the convention that the Bloch state is periodic with respect to the reciprocal lattice vectors.  We choose the Coulomb interaction screened by symmetric metal gate: $v(\bm{q}) = e^2\text{tanh}(|\bm{q}| d) / 2\epsilon_0 \epsilon |\bm{q}|$. Here, $e$ is the elementary charge; $d$ is the distance between tMoTe$_2$ and the metal gate; $\epsilon$ is relative permittivity; $\epsilon_0$ is the vacuum permittivity. The form factor $f_{n_1 n_2}(\bm{k}, \bm{q})$ is defined as
\begin{equation}
f_{n_1 n_2}(\bm{k}, \bm{q}) = \langle n_1 \bm{k} |\text{e}^{-\text{i}\bm{q} \cdot \hat{\bm{r}}}| n_2 \bm{k}+\bm{q} \rangle,
\end{equation}
where $|n, \bm{k}\rangle$ is a Bloch state.
To compute the form factor, we calculate the matrix element $\langle n_1 \bm{R}_1| \text{e}^{-\text{i}\bm{q} \cdot \hat{\bm{r}}} | n_2 \bm{R}_2 \rangle$, where $|n \bm{R}\rangle$ is the $n$th Wannier function sitting at the unit cell labeled by the lattice vector $\bm{R}$. $f_{n_1 n_2}(\bm{k}, \bm{q})$ can then be obtained by a Fourier transformation.

\begin{figure}
\centering
\includegraphics[width=0.988\columnwidth]{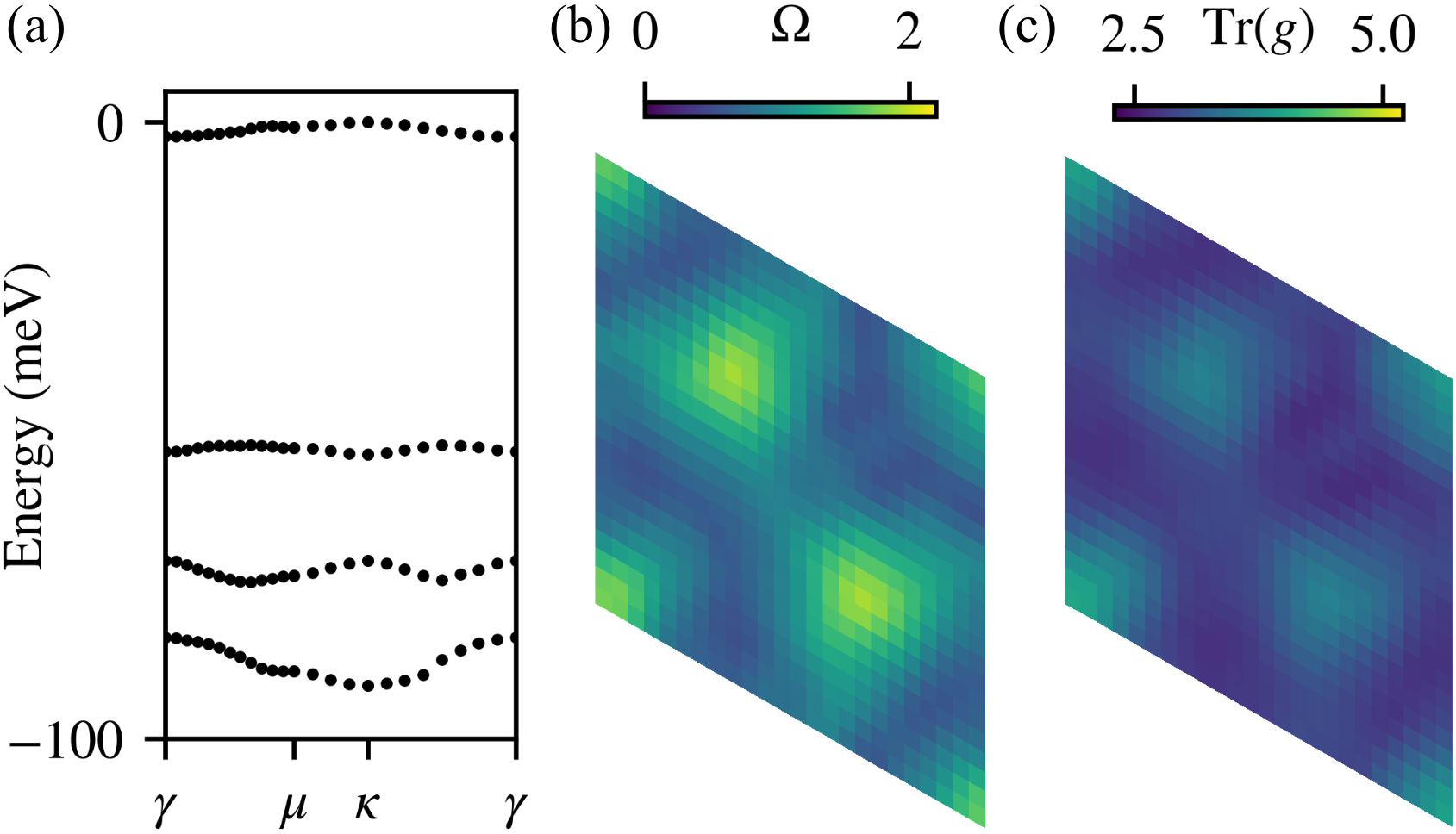}
\caption{
Hartree-Fock calculations with Wannier functions. (a) shows the Hartree-Fock quasiparticle bands for the $K$ valley. (b) and (c) show the distribution of the Berry curvature $\Omega$ and trace of the Fubini-Study metric tensor $\mathrm{tr}(g)$ in the Brillouin zone for the second moir\'e miniband after HF calculations. The color scale of (b) and (c) is the same as Fig.~\ref{fig:wannier}(b,c).
\label{fig:HF}}
\end{figure}

\textit{Signatures of non-Abelian states}.---One of the most fascinating features in higher LLs is the existence of non-Abelian states such as the MR state. The MR state can be thought of as superconducting paired CFLs~\cite{greiter1992paired,PhysRevB.61.10267}, and is known to be the exact ground state of a pure three-body short ranged interaction in the LLL~\cite{greiter1992paired,PhysRevLett.105.196801} or LLL-like Chern bands~\cite{PhysRevB.85.075116,PhysRevLett.108.126805,PhysRevB.85.075128,zhang2024moore}, or stabilized by the more realistic Coulomb interaction in the first LL. To explore whether this non-Abelian state can appear in tMoTe$_2$ at zero magnetic field, ED calculations are required. However, direct ED calculations at doping $\nu = -5/2$ (two and a half holes per moir\'e unit cell) are prohibitively demanding. Here, we carry out HF calculations at doping $\nu = -2$ to select the active orbitals.

The central object in HF calculations is the one-body reduced density matrix $\rho_{n_1 n_2}(\bm{k}_1, \bm{k}_2) = \langle \hat{a}^\dagger_{n_2 \bm{k_2}} \hat{a}_{n_1 \bm{k_1}}\rangle$. For an arbitrary $\rho$, a mean field decomposition of $\hat{H}_{\text{int}}$ gives rise to the HF interaction Hamiltonian $\hat{H}_{\text{HF}}[\rho]$. In DFT calculations, part of the electron-electron interaction has already been taken into account. To avoid double counting, we subtract the $\hat{H}_{\text{HF}}[\rho_0]$ from the total Hamiltonian:
\begin{equation}
    \hat{H} = \sum_{n, \bm{k}} \epsilon_{n\bm{k}} \hat{a}^\dagger_{n\bm{k}} \hat{a}_{n\bm{k}} + \hat{H}_{\text{int}} - \hat{H}_{\text{HF}}[\rho_0].\label{double_counting}
\end{equation}
Here, $\epsilon_{n\bm{k}}$ is the DFT band energy, and $\rho_0$ is the one-body reduced density matrix from the DFT calculations. HF calculations are carried out with $\hat{H}$ defined above. The subtraction of $\hat{H}_{\text{HF}}[\rho_0]$ ensures that when all valence bands are occupied, the HF calculations reproduce exactly the same band energies and Bloch states from DFT calculations.

The results of Hartree-Fock calculations are presented in Fig.~\ref{fig:HF}. Only frozen bands are included in the calculations. An enhanced gap between the first and second moir\'e valence bands can be observed. Crucially, after HF calculations, the fluctuations of $\Omega$ and $\mathrm{tr}(g)$ are significantly reduced by approximately 50\% (Table~\ref{tab:angles}), while $\chi$ slightly decreases (3.09 to 3.02). The improvement of quantum geometry enhances the analogy between the second moir\'e band and the first LL.

\begin{figure}
\centering
\includegraphics[width=0.994\columnwidth]{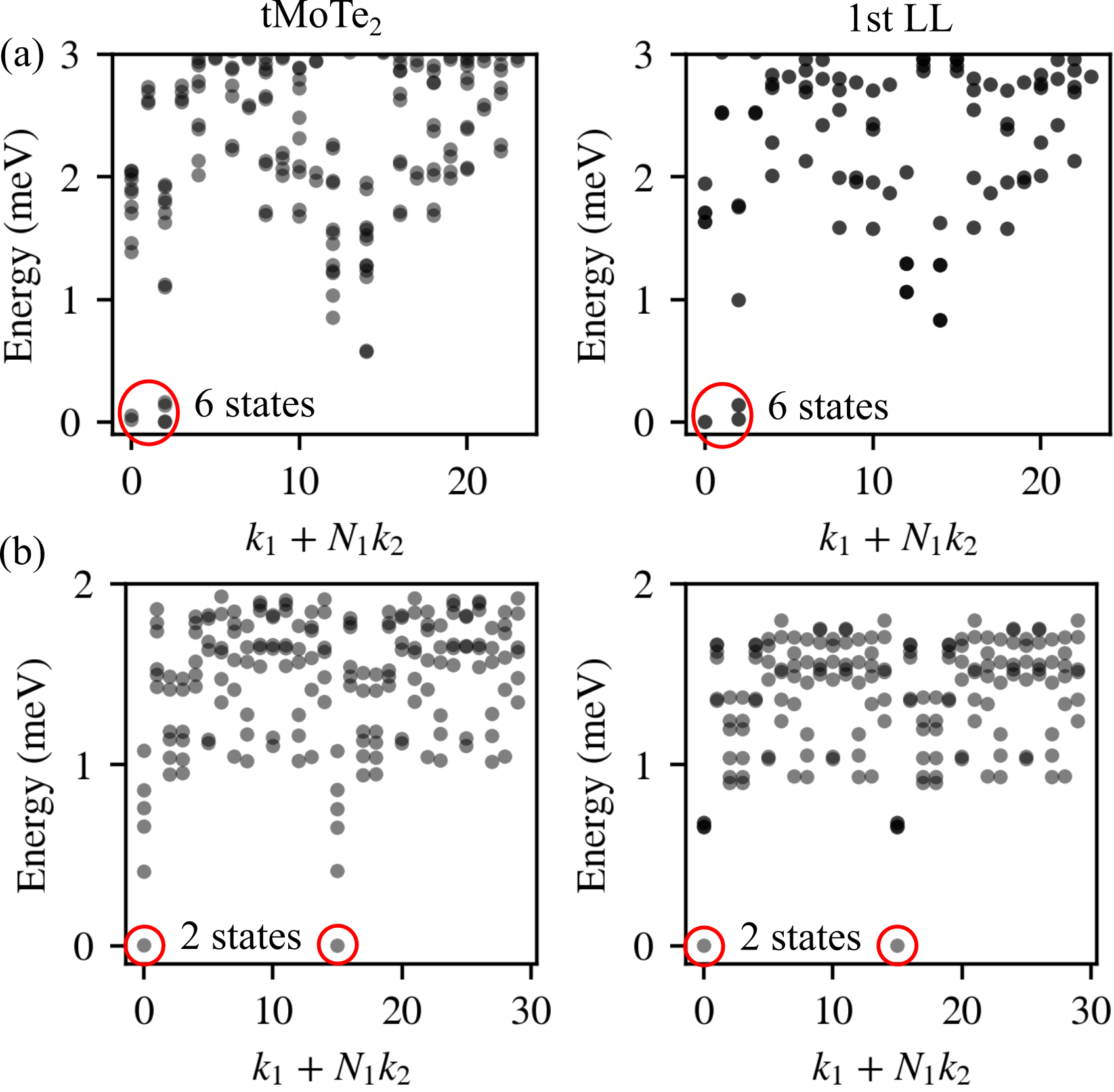}
\caption{Many-body spectrum of half-filled second moir\'e miniband (left) and half-filled first LL (right) on a $4 \times 6$ [(a)] and $5 \times 6$ [(b)] supercell with periodic boundary condition. The twist angle is $2.00^\circ$ for tMoTe$_2$. Parameters: $\epsilon=5$, $d=300$~\AA.
\label{fig:ED}}
\end{figure}

Focusing on the half-filled second moir\'e valence band, we carry out ED calculations on top of the HF calculations. To avoid double counting of the electron-electron interaction, we again utilize Eq.~(\ref{double_counting}) as the Hamiltonian in ED calculations. In this context, $\epsilon_{n \bm{k}}$ and $\rho_0$ in Eq.~(\ref{double_counting}) are band energies and one-body reduced density matrix from HF calculations. Previously, the double counting removing procedure for HF calculations on top of DFT calculations is heuristic. However, the same procedure, used for ED calculations on top of HF calculations, is exact. The sole purpose of HF calculations is to select relevant Bloch states as single particle orbitals for ED calculations.

On a $4 \times 3$ supercell with periodic boundary condition, our ED calculations, restricted to the second moir\'e valence bands from both valleys, show that fully valley polarized state is the ground state with parameters specified in the caption of Fig.~\ref{fig:ED}. Therefore, we further restrict the ED calculations to the second moir\'e valence band from the $K$ valley. The many-body spectrums are shown in Fig.~\ref{fig:ED} for $4 \times 6$ and $5 \times 6$ supercells with periodic boundary condition. The six-fold and two-fold ground state degeneracy with even and odd number of electrons is the hallmark of the non-Abelian states of MR type or its particle hole conjugate. In Fig.~\ref{fig:ED}, we also show the many-body spectrum of half-filled first LL with Coulomb interaction $v(\bm{q}) = e^2/2\epsilon_0\epsilon|\bm{q}|$. The LL system is put on the same torus as the corresponding tMoTe$_2$ system. The number of magnetic fluxes piercing the torus for the LL system is equal to the number of unit cells in the tMoTe$_2$ system. The striking similarity of the many-body spectrum between tMoTe$_2$ and the LL system is another strong indication of the non-Abelian states.

In the Supplemental Material~\cite{supplemental}, we present the many-body spectrum on a $4 \times 6$ supercell of half-filled second moir\'e miniband, but with bare DFT bands. The spectrum bears similarity to that of the half-filled first LL, but lacks the six-fold ground state degeneracy. Therefore, the improved quantum geometry from HF calculations is crucial for the non-Abelian states to appear.

We have also performed calculations for twist angles $1.89^\circ$ and $2.14^\circ$. The quantum geometry before and after HF calculations is presented in Table~\ref{tab:angles}. Evidence of non-Abelian states is also found at twist angle $2.14^\circ$, but with a smaller many-body gap. No non-Abelian states are found at $1.89^\circ$, for which the fluctuations of $\Omega$ and $\mathrm{tr}(g)$ are not significantly suppressed by HF calculations.

\begin{table}
\caption{
Many-body gaps (unit: meV) of non-Abelian state for tMoTe$_2$ and relevant band properties, including the integration of the quantum metric ($\chi$), the standard deviation of the Berry curvature ($\Delta\Omega$) and the quantum metric [$\Delta\mathrm{tr}(g)$], the band width of the second moir\'e miniband ($w$, unit: meV) at various twist angles ($\theta$). NaN indicates that no evidence of non-Abelian states has been found. The gap is identified on a $4 \times 6$ supercell and the parameters are the same as that specified in the caption of Fig.~\ref{fig:ED}. 
\label{tab:angles}
}
\begin{ruledtabular}
\begin{tabular}{c|cccc|cccc|c}
& \multicolumn{4}{c|}{Before HF} & \multicolumn{4}{c|}{After HF} & \\
$\theta$ & $\chi$ & $\Delta\Omega$ & $\Delta\mathrm{tr}(g)$ & $w$ & $\chi$ & $\Delta\Omega$ & $\Delta\mathrm{tr}(g)$ & $w$ & Gap \\
\hline
$1.89^\circ$ & 3.07 & 0.88 & 0.79 & 1.80 & 3.04 & 0.72 & 0.59 & 3.90 & NaN \\
$2.00^\circ$ & 3.09 & 0.51 & 0.67 & 0.95 & 3.02 & 0.32 & 0.27 & 1.59 & 0.41 \\
$2.14^\circ$ & 3.15 & 0.99 & 1.08 & 2.30 & 3.06 & 0.28 & 0.27 & 1.74 & 0.11 \\
\end{tabular}
\end{ruledtabular}
\end{table}

In LL systems, the MR (Pfaffian) state and its particle-hole (PH) conjugate anti-Pfaffian state are degenerate if LL mixing effects were ignored; the LL mixing provides a PH breaking effect and selects anti-Pfaffian over Pfaffian~\cite{PhysRevLett.119.026801}. Besides Pfaffian and anti-Pfaffian, an intrinsically PH symmetric topological order, PH-Pfaffian, was also proposed~\cite{Vishwanath_T_Pfaffian,PhysRevX.5.031027}. In our systems, the PH symmetry is explicitly broken by the dispersion and non-uniform quantum geometries. We leave more detailed examination of the precise nature of our ground state to the future work, which can be addressed by wave function overlap and entanglement spectrum analysis. Besides $\nu = -5/2$, we also perform ED calculations at other fillings. Specifically, at $\nu = -13/5$, the many-body spectrum for tMoTe$_2$ also resembles that of the first LL~\cite{supplemental}, for which the Read-Rezayi state~\cite{PhysRevB.59.8084} is supposed to be stabilized.

Compared to FCI states, signatures of non-Abelian states presented here are much weaker. It has been shown that the $\nu = -2/3$ Laughlin state exists in a wide range of twist angles in tMoTe$_2$~\cite{PhysRevLett.132.036501,PhysRevB.108.085117}. However, evidences of non-Abelian states are only found in a narrow window of twist angles in this work. In addition, the many-body gap in Fig.~\ref{fig:ED} is also several times smaller than that of the $\nu = -2/3$ FCI state at the same interaction strength. Finally, our choice of the dielectric constant $\epsilon$ gives rise to characteristic interaction strength that is several times larger than the band gap. Therefore, it should be critically evaluated whether the non-Abelian state is stable against band mixing in tMoTe$_2$, which was proved to be crucial in understanding FCI states in the same system at twist angle around $3.89^\circ$~\cite{PhysRevB.109.045147,xu2024maximally,abouelkomsan2024band}.

\begin{acknowledgments}
We acknowledge useful discussions with Yuchi He, Ying Ran, Lingnan Shen, Kai Sun and Zhao Liu.
The exact diagonalization study is supported by DOE Award No. DE-SC0012509.  The density-functional theory calculation is supported by the Center on Programmable Quantum Materials, an Energy Frontier Research Center funded by DOE BES under award DE-SC0019443. The machine learning of moir\'e structure is supported by the discovering AI@UW Initiative and by the National Science Foundation under Award DMR-2308979.  This work uses Microsoft Azure credits funded by discovering AI@UW Initiative.
\end{acknowledgments}

\textit{Note added.}---
We recently became aware of Refs.~[\onlinecite{reddy2024non,xu2024multiple,ahn2024first}]. Ref.~[\onlinecite{reddy2024non}] proposed the existence of non-Abelian states in free electron gas coupled to Skyrmion lattices. An updated version of Ref.~[\onlinecite{reddy2024non}] and Refs.~[\onlinecite{xu2024multiple,ahn2024first}] proposed the existence of non-Abelian states in tMoTe$_2$ based on continuum models fitted to DFT bands. {During the referee process, Ref.~\onlinecite{zhang2024universal} appeared and introduced the approach of constructing moir\'e continuum models from DFT bands without fitting. Up to 6 harmonics were included in the continuum model.}

\bibliography{main}

\end{document}


\title{Supplemental Material for ``Higher Landau-Level Analogs and Signatures of Non-Abelian States in Twisted Bilayer MoTe$_2$''}

\author{Chong Wang}
\thanks{These authors contribute equally to the work.}
\affiliation{Department of Materials Science and Engineering, University of Washington, Seattle, WA 98195, USA}
\author{Xiao-Wei Zhang}
\thanks{These authors contribute equally to the work.}
\affiliation{Department of Materials Science and Engineering, University of Washington, Seattle, WA 98195, USA}
\author{Xiaoyu Liu}
\affiliation{Department of Materials Science and Engineering, University of Washington, Seattle, WA 98195, USA}
\author{Jie Wang}
\affiliation{Department of Physics, Temple University, Philadelphia, Pennsylvania, 19122, USA}
\author{Ting Cao}
\email{tingcao@uw.edu}
\affiliation{Department of Materials Science and Engineering, University of Washington, Seattle, WA 98195, USA}
\author{Di Xiao}
\email{dixiao@uw.edu}
\affiliation{Department of Materials Science and Engineering, University of Washington, Seattle, WA 98195, USA}
\affiliation{Department of Physics, University of Washington, Seattle, WA 98195, USA}

\maketitle

\section{Details on Wannier Function Construction}

\begin{figure}
\centering
\includegraphics[width=0.638\textwidth]{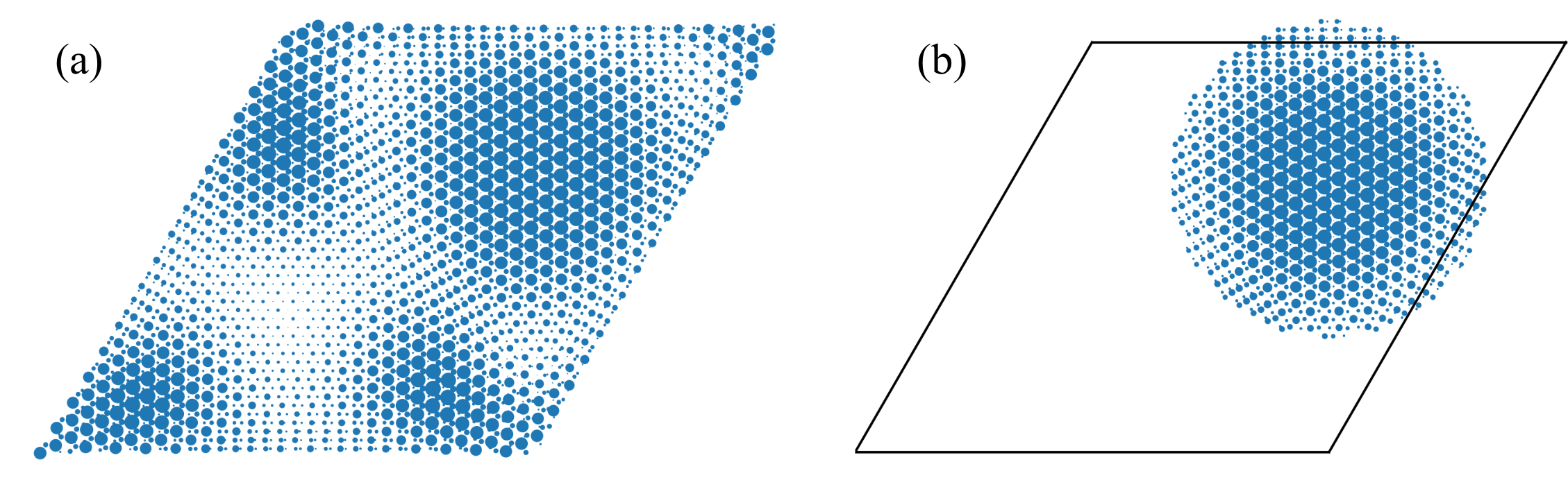}
\caption{
(a) Real space plot of the Bloch state for the third moir\'e miniband at $\kappa'$ point for tMoTe$_2$ at twist angle $2.00^\circ$. (b) The trial Wannier function constructed from the Bloch state in (a).
\label{fig:supp_wannier}}
\end{figure}

The common choices of the trial Wannier functions are atomic orbitals. However, for the purpose of describing moir\'e minibands, moir\'e-scale Wannier functions are needed. In this work, we choose the trial Wannier functions to be Bloch states at high symmetric $\bm{k}$ points, with a cutoff in real space around high symmetric stackings. There are three high symmetic stackings in tMoTe$_2$, namely the MX (Mo on top of Te), XM (Te on top of Mo) and MM (Mo on top of Mo) stackings. At some high symmetry $\bm{k}$ points, DFT wave functions for certain bands are localized at the MM stacking, prompting the adoption of orbitals at MM stacking in some literature. However, we find it more convenient to choose trial Wannier functions localized at the MX and XM stackings. The Wannier functions constructed are relatively extended and can reproduce the DFT wave function at the MM stacking. We note that due to the existence of flat Chern bands, the Wannier function is required to be relatively extended~\cite{chen2014impossibility}, and our decision of not choosing orbitals at the MM stacking does not significantly increase the size of the Wannier functions.

The six trial Wannier functions are constructed from bands $\{3, 3, 4, 4, 5, 5\}$, from $\bm{k}$-points $\{\kappa',\kappa,\kappa',\kappa,\kappa',\kappa\}$, centered at {MX, XM, MX, XM, MX, XM} stackings, respectively. The real space cutoff is one third of the lattice constant. One example of the Bloch states and the trial Wannier functions are shown in Fig.~\ref{fig:supp_wannier}.

\section{Details on Calculating Form Factors of Wannier Interpolation}

The core component to compute form factors for Bloch states is
$\bracket{w_n}{\mathe^{\mathi \tmmathbf{q} \cdot \hat{\tmmathbf{r}}}}{w_m}$,
where $\ket{w_n}$ is the $n$th Wannier function. The Wannier functions are in
turn linear combinations of local orbitals:
\begin{eqnarray}
  \ket{w_n} & = & \sum_{\tmmathbf{\tau}} \sum_{i \in
  \mathbb{O}_{\tmmathbf{\tau}}} c_{i\tmmathbf{\tau}n} \ket{\tmmathbf{\tau}i},
\end{eqnarray}
Local orbitals are classified according to atoms, which are labeled by their
positions $\tmmathbf{\tau}$. The list of orbitals for the atom sitting at
$\tmmathbf{\tau}$ is denoted as $\mathbb{O}_{\tmmathbf{\tau}}$. Therefore,
\begin{eqnarray}
  \bracket{w_n}{\mathe^{\mathi \tmmathbf{q} \cdot \hat{\tmmathbf{r}}}}{w_m} &
  = & \sum_{\{ \tmmathbf{\tau} \}} \sum_{\{ i_{\alpha} \in
  \mathbb{O}_{\tmmathbf{\tau}_{\alpha}} \}} c_{i \nocomma_1 \tmmathbf{\tau}_1
  n}^{\ast} c_{i_2 \tmmathbf{\tau}_2 \nocomma m} \bracket{\tmmathbf{\tau}_1
  i_1}{\mathe^{\mathi \tmmathbf{q} \cdot
  \hat{\tmmathbf{r}}}}{\tmmathbf{\tau}_2 i_2} .
\end{eqnarray}
We now make the approximation that $\bracket{\tmmathbf{\tau}_1
i_1}{\mathe^{\mathi \tmmathbf{q} \cdot \hat{\tmmathbf{r}}}}{\tmmathbf{\tau}_2
i_2} \neq 0$ only when $\tmmathbf{\tau}_1 \neq \tmmathbf{\tau}_2$. As a result
\begin{eqnarray}
  \bracket{w_n}{\mathe^{\mathi \tmmathbf{q} \cdot \hat{\tmmathbf{r}}}}{w_m} &
  \approx & \sum_{\tmmathbf{\tau}} \sum_{\{ i_{\alpha} \in
  \mathbb{O}_{\tmmathbf{\tau}} \}} c_{i \nocomma_1 \tmmathbf{\tau}n}^{\ast}
  c_{i_2 \tmmathbf{\tau} \nocomma m}
  \bracket{\tmmathbf{\tau}i_1}{\mathe^{\mathi \tmmathbf{q} \cdot
  \hat{\tmmathbf{r}}}}{\tmmathbf{\tau}i_2}\\
  & = & \sum_{\tmmathbf{\tau}} \sum_{\{ i_{\alpha} \in
  \mathbb{O}_{\tmmathbf{\tau}} \}} c_{i \nocomma_1 \tmmathbf{\tau}n}^{\ast}
  c_{i_2 \tmmathbf{\tau} \nocomma m} \mathe^{\mathi \tmmathbf{q} \cdot
  \tmmathbf{\tau}} \bracket{\tmmathbf{\tau}i_1}{\mathe^{\mathi \tmmathbf{q}
  \cdot (\hat{\tmmathbf{r}} -\tmmathbf{\tau})}}{\tmmathbf{\tau}i_2} .
\end{eqnarray}
Here, $\bracket{\tmmathbf{\tau}i_1}{\mathe^{\mathi \tmmathbf{q} \cdot
(\hat{\tmmathbf{r}} -\tmmathbf{\tau})}}{\tmmathbf{\tau}i_2}$ only depends on
$\tmmathbf{\tau}$ through the type of atom sitting at $\tmmathbf{\tau}$.

To carry out many-body calculations, it is necessary to introduce a cutoff for
$\tmmathbf{q}$, since otherwise the summation would be infinite. We assume
that $c_{i\tmmathbf{\tau}n}$ varies slowly in space, and therefore only small
$\tmmathbf{q}$ will give rise to significant $\bracket{w_n}{\mathe^{\mathi
\tmmathbf{q} \cdot \hat{\tmmathbf{r}}}}{w_m}$, introducing a natural cutoff
for $\tmmathbf{q}$. It is possible to further approximate
\begin{eqnarray}
  \bracket{\tmmathbf{\tau}i_1}{\mathe^{\mathi \tmmathbf{q} \cdot
  (\hat{\tmmathbf{r}} -\tmmathbf{\tau})}}{\tmmathbf{\tau}i_2} & \approx &
  \braket{\tmmathbf{\tau}i_1}{\tmmathbf{\tau}i_2}
\end{eqnarray}
for small $\tmmathbf{q}$ to simply reuse the overlap matrix from other
calculations.

\section{Details on Double Counting Removal Procedure}

We start with a general interacting Hamiltonian
\begin{eqnarray}
  H & = & \sum_{\alpha, \beta} T_{\alpha \beta} \hat{c}_{\alpha}^{\dagger} \hat{c}_{\beta}
  + \frac{1}{2} \sum_{\alpha, \beta, \gamma, \delta} V_{\alpha \beta \gamma
  \delta} \hat{c}_{\alpha}^{\dagger} \hat{c}_{\beta}^{\dagger} \hat{c}_{\gamma} \hat{c}_{\delta},
\end{eqnarray}
where $V_{\alpha \beta \gamma \delta} = V_{\beta \alpha \delta \gamma}$ is
assumed. Here, the single particle basis $\{ \alpha \}$ is supposed to be from
Hartree-Fock calculations. The Hartree-Fock Hamiltonian is generally
\begin{eqnarray}
  H_{\mathrm{HF}} & = & \sum_{\alpha, \beta} T_{\alpha \beta}
  \hat{c}_{\alpha}^{\dagger} \hat{c}_{\beta} + \sum_{\alpha, \beta, \gamma, \delta}
  V_{\alpha \beta \gamma \delta} \left[ \average{\hat{c}_{\alpha}^{\dagger}
  \hat{c}_{\delta}} \hat{c}_{\beta}^{\dagger} \hat{c}_{\gamma} - \average{\hat{c}_{\alpha}^{\dagger}
  \hat{c}_{\gamma}} \hat{c}_{\beta}^{\dagger} \hat{c}_{\delta} \right],
\end{eqnarray}
but since $\{ \alpha \}$ are eigenstates, $H_{\mathrm{HF}}$ is diagonal
\begin{eqnarray}
  T_{\alpha \beta} + \sum_{\gamma} n_{\gamma} (V_{\gamma \alpha \beta \gamma}
  - V_{\gamma \alpha \gamma \beta}) & = & \varepsilon_{\alpha} \delta_{\alpha
  \beta},
\end{eqnarray}
where $n_{\alpha}$ is the occupation number of the state $\alpha$ in the
Hartree-Fock calculations, and $\varepsilon_{\alpha}$ is the Hartree-Fock
quasiparticle energy. We currently work at zero temperature, such that
$n_{\alpha}$ is either $1$ or $0$.

To go beyond the Hartree-Fock calculation and to allow number of electrons to
be different from what is set in the Hartree-Fock calculations, we now divide
the quasiparticle states into several groups. The first group includes
quasiparticle states with quasiparticle energies much lower than the Fermi
energy. These states are called frozen states, and all other states are called
free states. To facilitate further reasoning, we defined another group of
states, called melted states. Melted states includes quasiparticle states with
occupation number $1$ in the Hartree-Fock calculations but is now allowed to
change. Melted states is a subset of free states. Melted states can be empty,
for which all the occupied quasiparticle states in the Hartree-Fock
calculations are frozen.

The Hilbert space defined above is called the reduced Hilbert space. The
simplification is that frozen states are always occupied, which greatly
reduces the number of electrons that need to be dealt with. Out ultimate
purpose is to derive an effective Hamiltonian in the reduced Hilbert space.

We first look at the kinetic part $\sum_{\alpha, \beta} T_{\alpha \beta}
\hat{c}_{\alpha}^{\dagger} \hat{c}_{\beta}$. If both $\alpha$ and $\beta$ belongs to the
frozen states, the term is a constant in the restricted Hilbert space. If one
of $\{ \alpha, \beta \}$ belongs to the frozen states and the other one
belongs to the free states, the term is zero in the restricted Hilbert space.
Therefore, the effective kinetic Hamiltonian is
\begin{eqnarray}
  H_{\mathrm{eff}}^{\mathrm{kin}} & = & \sum_{\alpha, \beta \in \mathrm{free}}
  T_{\alpha \beta} \hat{c}_{\alpha}^{\dagger} \hat{c}_{\beta} .
\end{eqnarray}
With the same logic, the effective interaction Hamiltonian is
\begin{eqnarray}
  H_{\mathrm{eff}}^{\mathrm{int}} & = & \frac{1}{2} \sum_{\alpha, \beta, \gamma,
  \delta \in \mathrm{free}} V_{\alpha \beta \gamma \delta} \hat{c}_{\alpha}^{\dagger}
  \hat{c}_{\beta}^{\dagger} \hat{c}_{\gamma} \hat{c}_{\delta} + \sum_{\alpha, \beta \in
  \mathrm{free} ; \gamma \in \mathrm{frozen}} (V_{\gamma \alpha \beta \gamma} -
  V_{\gamma \alpha \gamma \beta}) \hat{c}_{\alpha}^{\dagger} \hat{c}_{\beta} .
\end{eqnarray}
Together, the effective Hamiltonian is
\begin{eqnarray}
  H_{\mathrm{eff}} & = & \sum_{\alpha, \beta \in \mathrm{free}} \left[ T_{\alpha
  \beta} + \sum_{\gamma \in \mathrm{frozen}} (V_{\gamma \alpha \beta \gamma} -
  V_{\gamma \alpha \gamma \beta}) \right] \hat{c}_{\alpha}^{\dagger} \hat{c}_{\beta} +
  \frac{1}{2} \sum_{\alpha, \beta, \gamma, \delta \in \mathrm{free}} V_{\alpha
  \beta \gamma \delta} \hat{c}_{\alpha}^{\dagger} \hat{c}_{\beta}^{\dagger} \hat{c}_{\gamma}
  \hat{c}_{\delta}\\
  & = & \sum_{\alpha \in \mathrm{free}} \varepsilon_{\alpha}
  \hat{c}_{\alpha}^{\dagger} \hat{c}_{\alpha} + \frac{1}{2} \sum_{\alpha, \beta, \gamma,
  \delta \in \mathrm{free}} V_{\alpha \beta \gamma \delta} \hat{c}_{\alpha}^{\dagger}
  \hat{c}_{\beta}^{\dagger} \hat{c}_{\gamma} \hat{c}_{\delta} - \sum_{\alpha, \beta \in
  \mathrm{free} ; \gamma \in \mathrm{melted}} (V_{\gamma \alpha \beta \gamma} -
  V_{\gamma \alpha \gamma \beta}) \hat{c}_{\alpha}^{\dagger} \hat{c}_{\beta},
\end{eqnarray}
where the last term is the double counting term to be removed.

\begin{figure}
\centering
\includegraphics[width=0.806\textwidth]{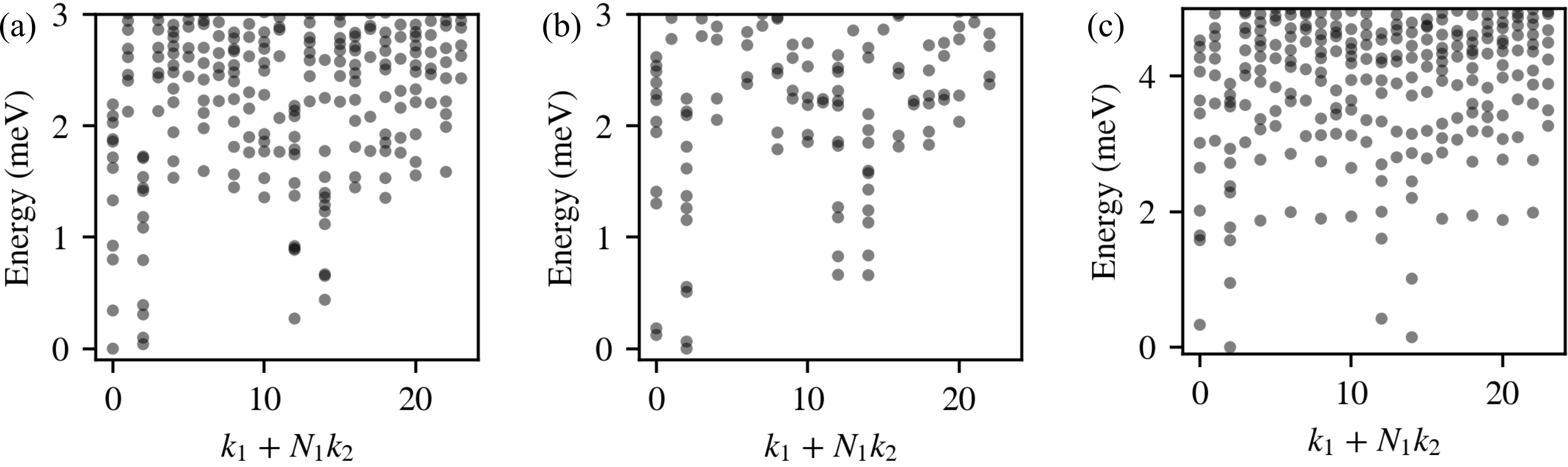}
\caption{
Many-body spectrum on a $4 \times 6$ supercell for tMoTe$_2$ at twist angle $1.89^\circ$ [(a)] and $2.14^\circ$ [(b)]. The ED calculations for (a) and (b) are performed after HF calculations. (c) shows the many-body spectrum of tMoTe$_2$ at twist angle $2.00^\circ$, but with bare DFT bands. For all calculations, the Hilbert space is restricted to the second moir\'e miniband, which is half-filled. The first moir\'e miniband is empty. The parameters are the same as that used in the main text.
\label{fig:supp_ED}}
\end{figure}

\section{Additional Many-body Calculations for twisted bilayer MoTe$_2$}

\subsection{Many-body spectra supporting the main text}

Additional many-body spectrums for tMoTe$_2$ at various twist angles are shown in Fig.~\ref{fig:supp_ED}. They show (1) absence of non-Abelian states at twist angle $1.89^\circ$ with ED on top of HF calculations; (2)
evidence of non-Abelian states at twist angle $2.14^\circ$ with ED on top of HF calculations; (3) absence of non-Abelian states at twist angle $2.00^\circ$ with ED on top of DFT calculations.

We have also performed ED calculations on a $4 \times 3$ supercell, the spin gap is 4.32~meV for tMoTe$_2$ at the twist angle $2.00^\circ$. The ED calculation is performed on top of HF calculations and the filling factor is $\nu = -5/2$. Here, the definition of spin gap is defined as the energy difference between the lowest non-fully-valley-polarized state and the full-valley-polarized ground state.

\subsection{Convergence test for number of bands in Hartree-Fock calculations}

\begin{figure}
\centering
\includegraphics[width=0.806\textwidth]{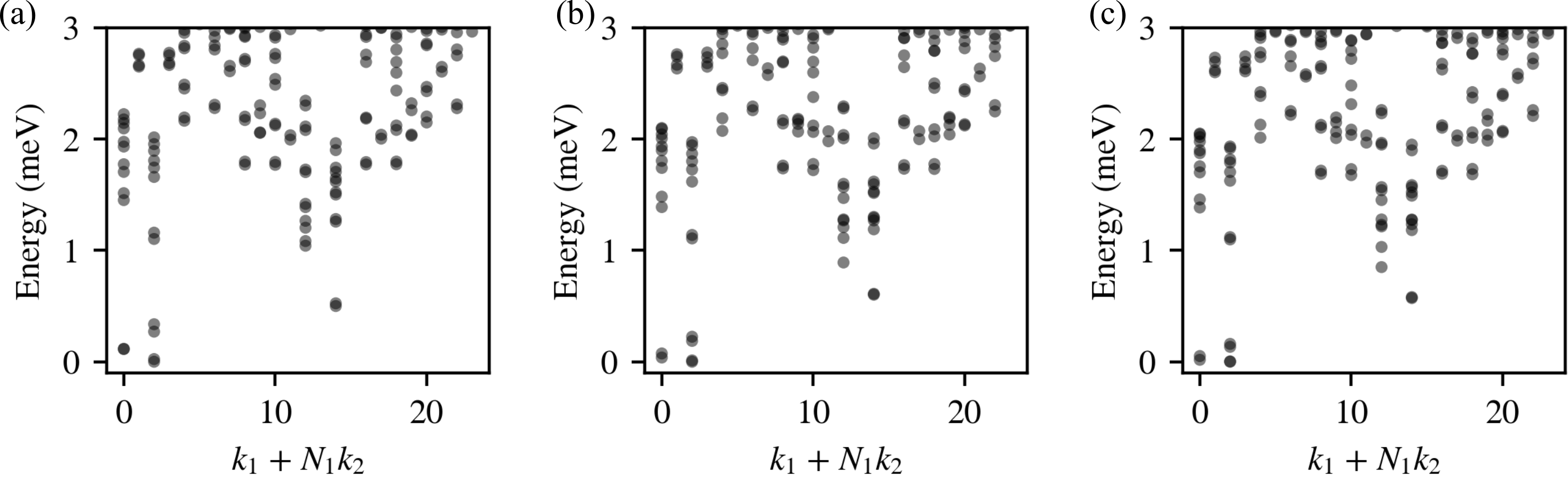}
\caption{
Many-body spectra on a $4 \times 6$ supercell for tMoTe$_2$ at twist angle 2.00$^\circ$. Top two [(a)], three [(b)], four [(c)] bands are included in the Hartree-Fock calculations. The parameters are the same as that used in the main text.
\label{fig:supp_varying_HF_bands}}
\end{figure}

Hartree-Fock calculations mix Bloch states from different DFT bands. To test whether enough bands have been taken into account in the HF calculations, we include two, three and four bands in the HF calculations and perform ED on top of the HF bands for tMoTe$_2$ at twist angle $2^\circ$. The many-body spectra, presented in Fig.~\ref{fig:supp_varying_HF_bands}, show minimal changes between three-band and four-band calculations, confirming the convergence of the results presented in the main text. Since ED calculations are restricted to the second moir\'e valence band, the above results also show that the second DFT moir\'e band mainly receives contribution from the first and the third DFT moir\'e valence bands.

\subsection{Influence of dielectric constant and the gate-sample distance}

\begin{figure}
\centering
\includegraphics[width=0.323\textwidth]{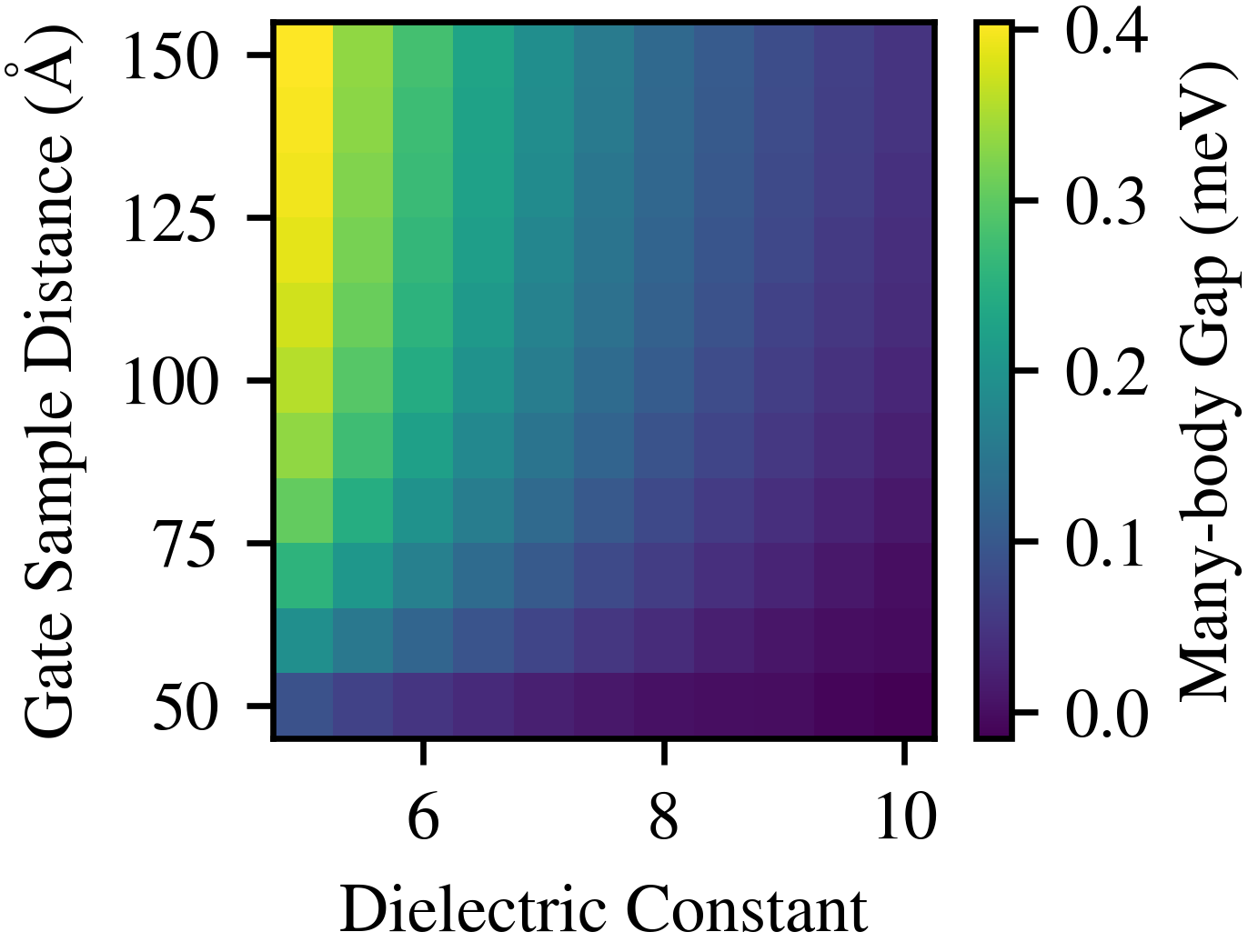}
\caption{
Many-body gap as a function of dielectric constant and the gate-sample distance. In this figure, the system is in the non-Abelian state for all combinations of parameters. The many-body gap is identified on a $4 \times 6$ supercell for tMoTe$_2$ at twist angle 2.00$^\circ$.
\label{fig:supp_epsilon_d}}
\end{figure}

We investigate the stability of the non-Abelian state with respect to screening, controlled by the dielectric constant and the gate-sample distance. In Fig.~\ref{fig:supp_epsilon_d}, we present the many-body gap as a function of the above two parameters. The range of the gate sample distance is chosen such that it is comparable to the moir\'e lattice constant ($\sim 100$~\AA~at twist angle $2^\circ$). The non-Abelian state is suppressed by both types of screening.

\subsection{Exact diagonalization calculations with more system sizes}

\begin{figure}
\centering
\includegraphics[width=0.998\textwidth]{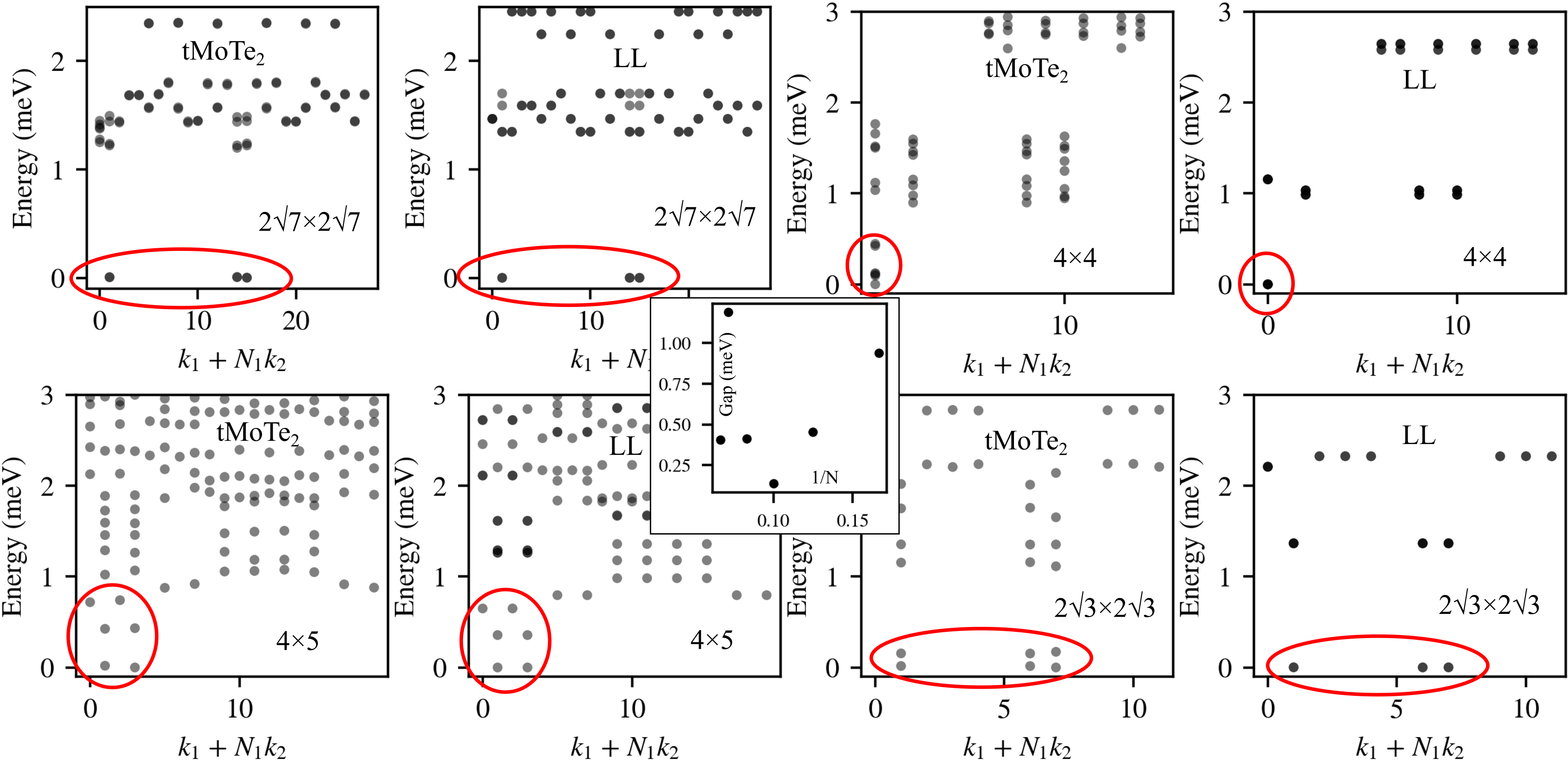}
\caption{
Exact diagonalization spectrum for half-filled tMoTe$_2$ with various system sizes. The red circles each enclose 6 ground states. The inset shows the many-body gap as a function of the inverse of the particle number, including the four systems sizes shown here and another two shown in the main text. \label{fig:supp_system_size}}
\end{figure}

To show that our calculations are robust with respect to system sizes. We perform exact diagonalization on four additional geometry and present the results in Fig.~\ref{fig:supp_system_size}. All calculations consistently exhibit six ground states, as expected for systems with an even number of electrons. Furthermore, the spectra for tMoTe$_2$ closely resemble those of the first Landau level.

\section{Signatures of Read-Rezayi state in twisted bilayer MoTe$_2$}

\begin{figure}
\centering
\includegraphics[width=0.540\textwidth]{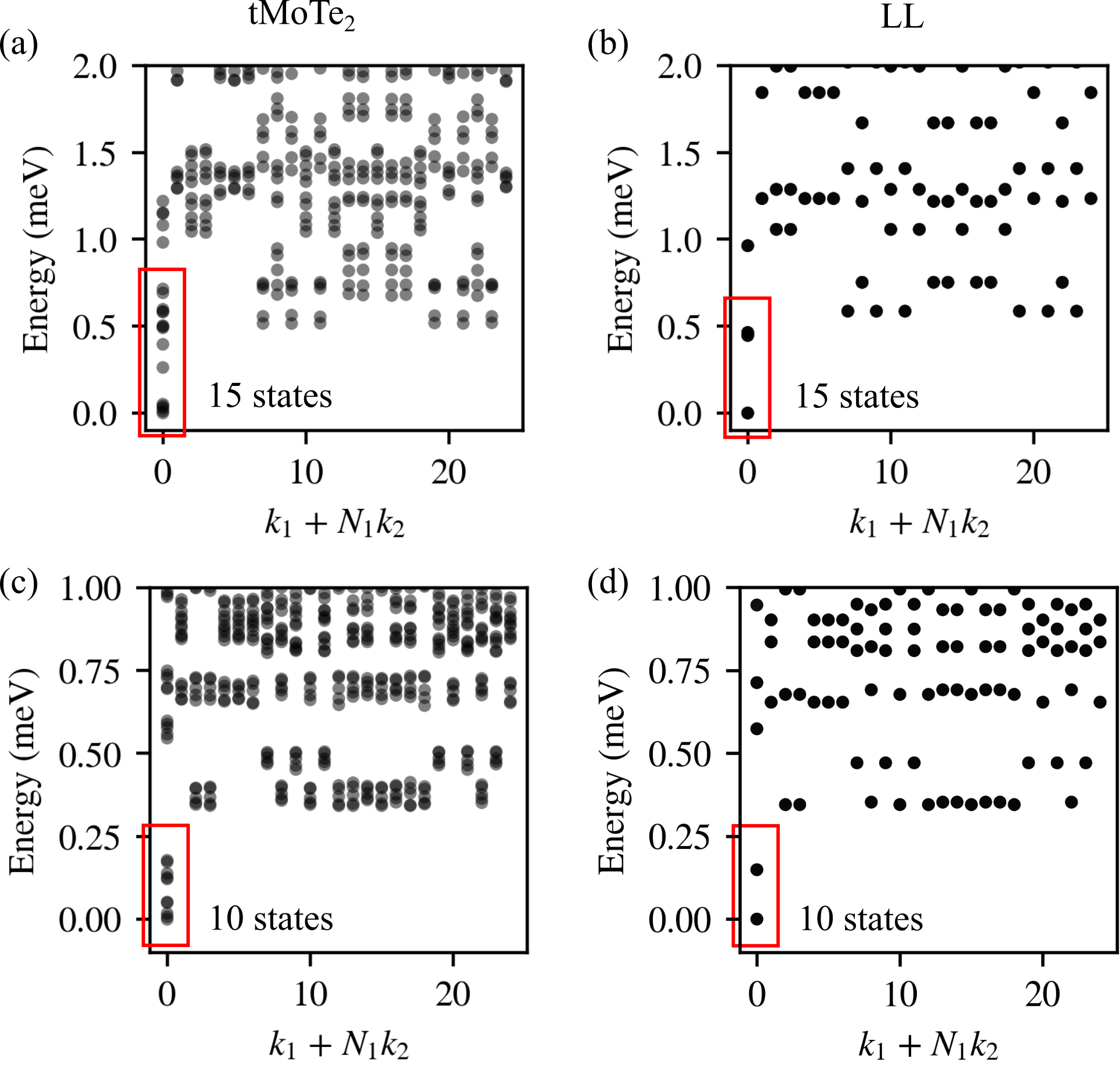}
\caption{
Many-body spectra for tMoTe$_2$ (left) and first LL (right) on a $5 \times 5$ supercell. The filling for tMoTe$_2$ is $\nu = -13/5$, and the twist angle is $2.00^\circ$. For (a) and (b), the interaction parameters are the same used in the main text. For (c) and (d), additional screening has been included in the calculations.
\label{fig:supp_read_rezayi}}
\end{figure}

Given that the second moir\'e miniband resembles the first LL, it is natural to ask whether the Read-Rezayi state~\cite{PhysRevB.59.8084} can be realized in tMoTe$_2$. In Fig.~\ref{fig:supp_read_rezayi}, we present ED spectra for tMoTe$_2$ at hole filling $\nu = -13/5$ and first LL at electron filling $\nu = 2/5$. The LL system is supposed to be in the Read-Rezayi state.

The ED evidence for the Read-Rezayi state is a 10-fold ground state degeneracy at the sector of zero total momentum for a $5 \times 5$ supercell. However, even for the first LL, no obvious gap can be observed between the 10th and the 11th state [Fig.~\ref{fig:supp_read_rezayi}(b)], possibly due to finite size effects. The 10-fold degeneracy can be revealed by adding additional screening to the interaction [Fig.~\ref{fig:supp_read_rezayi}(d)]:
\begin{equation}
    v(\bm{q}) = \frac{e^2}{2\epsilon_0 \epsilon |\bm{q}| [1 + 6\mathrm{tanh}(|\bm{q}|^2 l_B^2/2) / |\bm{q}|l_B]},
\end{equation}
where $l_B$ is the magnetic length. The above form of screening comes from virtual excitations of electron-hole pairs in LLs in bilayer graphene~\cite{PhysRevLett.112.046602}. It is not completely reasonable to use this type of screening in the context of tMoTe$_2$. Nevertheless, the ED spectrum for tMoTe$_2$ [Fig.~\ref{fig:supp_read_rezayi}(a,c)] resembles that of the LL system [Fig.~\ref{fig:supp_read_rezayi}(b,d)] regardless of the choice of $v(\bm{q})$. Therefore, it is possible that Read-Rezayi state can be realized in tMoTe$_2$ at zero magnetic field.

\section{Particle entanglement spectrum for nonabelian states in twisted bilayer MoTe$_2$}

Generally, to calculate entanglement spectrum, the Hilbert space $\mathcal{H}$
is needed to be decomposed into the tensor product of two different Hilbert
spaces:
\begin{eqnarray}
  \mathcal{H} & = & \mathcal{H}_A \otimes \mathcal{H}_B .
\end{eqnarray}
For example, a spin chain can be split into two, and the original Hilbert
space is the tensor product of the Hilbert spaces associated with the two
parts of the spin chain.

However, Hilbert space decomposition by splitting the particles is subtle. To
construct such a decomposition, it is necessary to enlarge the Hilbert space.
The original Hilbert space of wave functions containing $N$ particles has to
be symmetric or antisymmetric with respect to particle permutation. For $N =
N_A + N_B$, we enlarge the Hilbert space to be spanned by $\varphi_{\alpha}
(\tmmathbf{r}_1, \tmmathbf{r}_2, \ldots \tmmathbf{r}_{N_A}) \varphi_{\beta}
(\tmmathbf{r}_{N_A + 1}, \tmmathbf{r}_{N_A + 2}, \ldots \tmmathbf{r}_{N_A +
N_B})$, where $\alpha$ ($\beta$) labels a basis where $N_A$ ($N_B$) particles
occupy the original orbitals. Notice that for $\varphi_{\alpha}
(\tmmathbf{r}_1, \tmmathbf{r}_2, \ldots \tmmathbf{r}_{N_A})$ and
$\varphi_{\beta} (\tmmathbf{r}_{N_A + 1}, \tmmathbf{r}_{N_A + 2}, \ldots
\tmmathbf{r}_{N_A + N_B})$, we still require particle permutation symmetry or
antisymmetry. The above enlargement of the Hilbert space corresponds to
dividing the particles into two species, with $N_A$ particles belonging to one
species and $N_B$ particles belonging to the other species. Particles from
different species are distinguishable.

As an example, we work with four fermions, with $N = 4$ and $N_A = N_B = 2$. A
common basis for the original (not enlarged) Hilbert space is the Slater
determinant, which can be decomposed with generalized Laplace expansion
\begin{eqnarray*}
  \left| \begin{array}{cccc}
    \phi_1 (\tmmathbf{r}_1) & \phi_2 (\tmmathbf{r}_1) & \phi_3
    (\tmmathbf{r}_1) & \phi_4 (\tmmathbf{r}_1)\\
    \phi_1 (\tmmathbf{r}_2) & \phi_2 (\tmmathbf{r}_2) & \phi_3
    (\tmmathbf{r}_2) & \phi_4 (\tmmathbf{r}_2)\\
    \phi_1 (\tmmathbf{r}_3) & \phi_2 (\tmmathbf{r}_3) & \phi_3
    (\tmmathbf{r}_3) & \phi_4 (\tmmathbf{r}_3)\\
    \phi_1 (\tmmathbf{r}_4) & \phi_2 (\tmmathbf{r}_4) & \phi_3
    (\tmmathbf{r}_4) & \phi_4 (\tmmathbf{r}_4)
  \end{array} \right| & = & \left| \begin{array}{cc}
    \phi_1 (\tmmathbf{r}_1) & \phi_2 (\tmmathbf{r}_1)\\
    \phi_1 (\tmmathbf{r}_2) & \phi_2 (\tmmathbf{r}_2)
  \end{array} \right| \left| \begin{array}{cc}
    \phi_3 (\tmmathbf{r}_3) & \phi_4 (\tmmathbf{r}_3)\\
    \phi_3 (\tmmathbf{r}_4) & \phi_4 (\tmmathbf{r}_4)
  \end{array} \right| - \left| \begin{array}{cc}
    \phi_1 (\tmmathbf{r}_1) & \phi_3 (\tmmathbf{r}_1)\\
    \phi_1 (\tmmathbf{r}_2) & \phi_3 (\tmmathbf{r}_2)
  \end{array} \right| \left| \begin{array}{cc}
    \phi_2 (\tmmathbf{r}_3) & \phi_4 (\tmmathbf{r}_3)\\
    \phi_2 (\tmmathbf{r}_4) & \phi_4 (\tmmathbf{r}_4)
  \end{array} \right| \\  + & & \left| \begin{array}{cc}
    \phi_1 (\tmmathbf{r}_1) & \phi_4 (\tmmathbf{r}_1)\\
    \phi_1 (\tmmathbf{r}_2) & \phi_4 (\tmmathbf{r}_2)
  \end{array} \right| \left| \begin{array}{cc}
    \phi_2 (\tmmathbf{r}_3) & \phi_3 (\tmmathbf{r}_3)\\
    \phi_2 (\tmmathbf{r}_4) & \phi_3 (\tmmathbf{r}_4)
  \end{array} \right| + ...
\end{eqnarray*}
The above equation directly specifies how a Slater determinant should be
decomposed in the enlarged Hilbert space.

\begin{figure}
\centering
\includegraphics[width=0.518\textwidth]{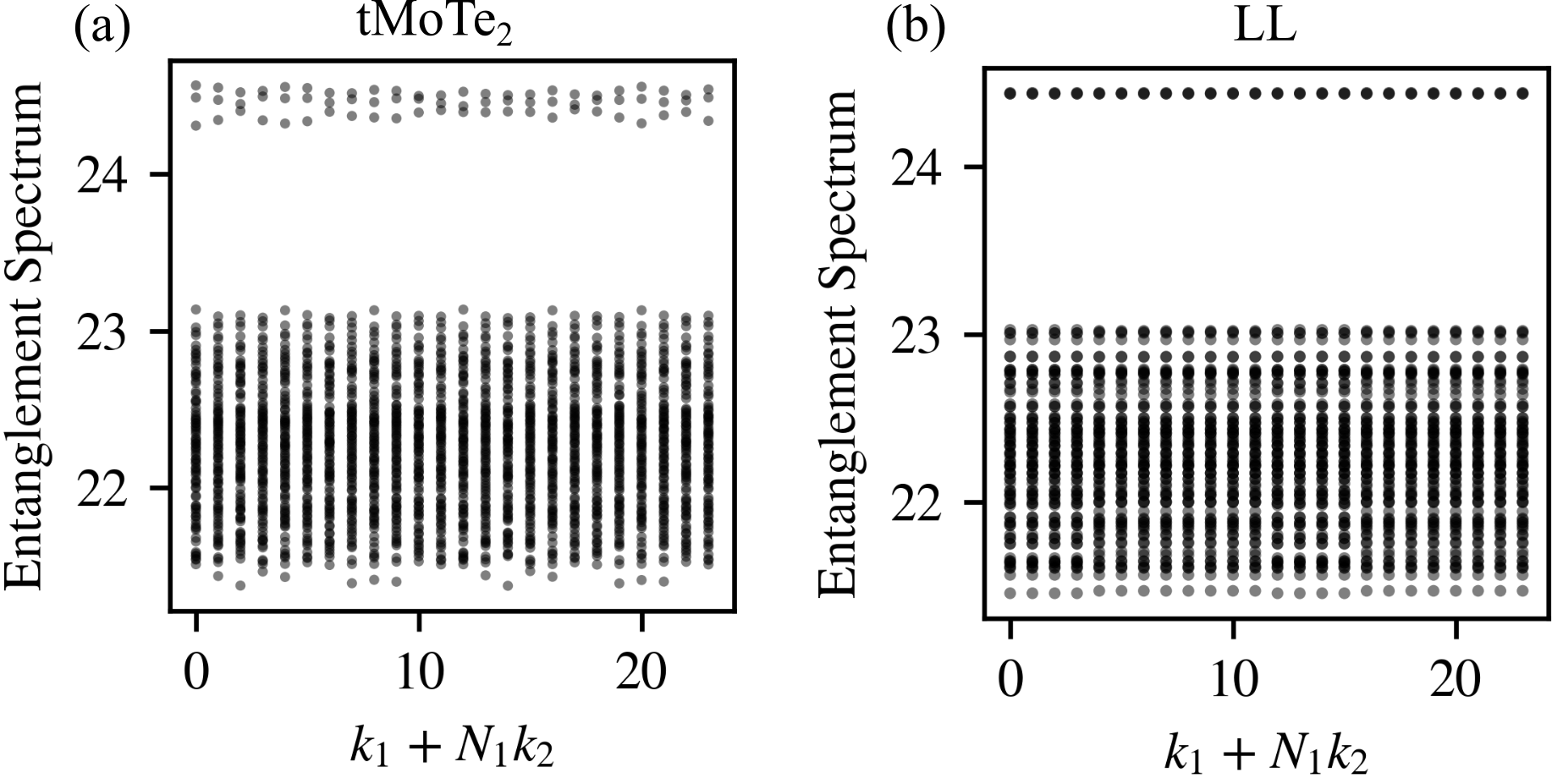}
\caption{
Particle entanglement spectrum for half-filled tMoTe$_2$ (left) and first LL (right) on a $4 \times 6$ supercell. To calculate the spectrum, the system is broken into two subsystems, with one of the subsystem having 3 particles. The number of levels below the spectrum gap for both (a) and (b) is 1952.
\label{fig:supp_entanglement_spectrum}}
\end{figure}

In Fig.~\ref{fig:supp_entanglement_spectrum}, we show the particle entanglement spectrum for half-filled tMoTe$_2$ and first LL. Both show 1952 levels below the spectrum gap, which corresponds to the number of quasihole excitations in the two systems.

\bibliography{main}